\documentclass[12pt,aps,prb,preprint]{revtex4}   

\usepackage{amsmath}    
\usepackage{amsfonts}  
\usepackage{amssymb}

\begin{document}

\title{Visual Observability of the Cassiopeia A Supernova}

\author{J. A. MORGAN}
\affiliation{The Aerospace Corporation~\cite{affil}}


\begin{abstract}
It is generally believed that the explosion which gave birth to the Cassiopeia A supernova remmant 
resulted from core collapse of a hydrogen-deficient star.  A  
progenitor that has lost all its hydrogen envelope and part of its helium envelope 
would lead to an explosion with the optical properties of a Type Ic 
supernova.  There is evidence, if not general agreement, that Flamsteed observed the Cas A supernova 
as a 
sixth magnitude object in August, 1680.  If an explosion with a typical SNIc light curve at the 
position and distance of Cas A attained maximum luminosity during the winter of 1679-1680, 
it would at that time have been poorly situated for visual observation, as its
upper culmination would have taken place during daylight,  
while in August, between 170-200 days after peak luminosity, it would have been a sixth magnitude
star.

\end{abstract}

\maketitle


A persistent
enigma regarding the origins of the Cassiopeia A supernova remnant is the dearth of 
contemporary accounts of a bright new star
at the time of the outburst as estimated
from the remnant expansion age.  Kinematical studies of the high proper-motion clouds known as 
fast-moving knots point to an explosion in $1672 \pm 18$.~\cite{Fesen2006}  Reed et 
al.~\cite{ReedEtAl1995} find a distance to the Cas A remnant of $d=3.4^{+0.3}_{-0.1}$ kpc, 
obtained from radial velocities and proper motions of a large number of fast-moving knots.  Their 
value agrees with that given earlier by Shklovski.~\cite{Shklovski1968}
At this distance, assuming normal extinction of one magnitude/kpc, a Type Ia supernova 
explosion should have had a peak visual magnitude of about -3, comparable to Venus.
On the other hand, van den Bergh and 
Dodd~\cite{vdBDodd1970} estimate that the Cas A supernova 
might not have
exceeded $+2$ magnitudes at maximum light, and, noting the frequency with which both Oriental and 
European astronomers
failed to record nova outbursts during this period, conclude that the absence of contemporary 
observations is unsurprising.
The only seventeenth century record of a
star near the location of Cas A is a single report by Flamsteed, who observed the sixth
magnitude star 3 Cassiopeiae on 1680 August 16 (Julian).~\cite{Ash80}  This star has not been 
observed subsequently, and the 
validity of Flamsteed's observation has been contested.~\cite{StephGreen2002}
Setting aside the question of whether or not Flamsteed did observe a star near the present-day
location of the Cas A remnant, it seems worth trying to construct a plausible sequence of events
that explains both the disputed observation of 3 Cassiopeiae by Flamsteed and 
the lack of other contemporary reports of a new star.  This note examines the possibility that
Cas A was a Type Ic supernova (at least, as to light curve) that 
attained peak luminosity in the winter of 1679-1680, most probably 
in February or March, 1680.  Flamsteed's observation of 3 Cassiopeiae, therefore, would have been
made some six months after maximum light of the Cas A supernova.
As will become apparent, the acccount to be presented does not tell a complete tale, and is at points
necessarily conjectural. 

It has long been thought that the Cas A supernova resulted from core collapse of a hydrogen-deficient
star.  The discovery~\cite{Tannanbaum1999} of a compact X-ray source, 
interpreted as a neutron star,~\cite{Chakrabarty2001} near
the inferred expansion center of the fast-moving knots and the observation of 
significant $^{44}Ti$ 
$\gamma$-emission~\cite{Iyudin1994} provide direct evidence for a core-collapse 
origin. Very little hydrogen emission has been 
observed in the fast-moving knots.  If we posit a progenitor that has been stripped of its hydrogen
envelope
altogether, the resulting explosion would be an SNIb or
SNIc supernova.~\cite{footnote2}

There are difficulties with assuming that Cas A supernova was of Type Ic.
To produce an SNIc light curve 
requires a progenitor that is (1) compact and (2) of low mass.  No direct evidence bearing 
on the radius of the Cas A progenitor at the time of the explosion is available to us, but from the 
remnant one might hope to learn somewhat concerning its mass.  
Estimates of the Cas A progenitor mass at the time of the explosion vary widely.  
One-dimensional hydrodynamic
calculations of the explosion of helium stars that have undergone significant mass loss 
reproduce the light curve of an SNIc outburst with a progenitor mass between 
$2.3 M_{\odot}-3.6 M_{\odot}$~\cite{WoosleyEtAl1995,Ensman1988}  This mass range comports 
awkwardly with estimates of the supernova ejecta mass of order 
$4 M_{\odot}$.~\cite{McKeeTrulove1995,VinkEtAl1996}
Measured abundance ratios of 
nucleosynthetic
products appear consistent with a mass of ca. $12 M_{\odot}$,~\cite{Willingale2001} while  
a recent three-dimensional numerical study of the evolution and explosion of Cas A finds that a WN 
Wolf-Rayet star 
of mass between 
$4M_{\odot}-6 M_{\odot}$ best fits combined constraints posed by nucleosynthesis, ejecta mass, 
and compact remnant.~\cite{Young2006}
Explosion of a Wolf-Rayet star of appreciable mass would presumably result in an SNIb outburst.  
There is also 
evidence 
that the Cas A progenitor may have had a very thin layer of hydrogen$^{17}$.
However, the light curve of the peculiar SNII SN1993J
also fits the requirements of the scenario presented here.

Cas A lies near the Galactic plane at the low latitude of 
$-2.1^\circ$, and extinction appears to be nonuniform across the remnant. 
Hurford and Fesen~\cite{Hurford1996} have determined from
$[SII]$ line ratios that the extinction to five fast-moving knots lies in the range $A_{v}=4.6-5.4$ 
within errors estimated as $\pm (0.25-0.45)$. 
Using (less certain) Balmer $H\alpha/H\beta$ ratios, they also find 
$A_{v} \le 5.3^{+0.9}_{-0.9}$ and $\le 6.2^{+0.9}_{-0.9}$ 
to two of the low proper-motion quasistationary flocculi. 
These values are
about one magnitude greater than the value $A_{v}=4.3$ previously found by 
Searle,~\cite{Searle1971} and suggest that the extinction varies across the remnant by as much as a 
magnitude. 
The distance modulus for Cas A is $12.7^{+0.2}_{-0.1}$.
If we provisionally accept Flamsteed's observation of a sixth magnitude star near the location of the 
present-day
Cas A supernova remnant, and estimate the visual extinction to that remnant to be
$A_{v}=5^{+0.45}_{-0.45}$,
then the absolute magnitude of the supernova was
approximately 
$M_{V}=-11.7^{+0.5}_{-0.4}$ 
in August, 1680.  This value is characteristic of late 
$> 200$ day behavior of a supernova
light curve, rather than an early peak value of approximately $M_{V}=-16$ to $-18$, suggesting 
that the initial outburst could have taken place as much as $200$ days earlier than 1680 August 16.  

At any time
between, roughly, New Year's 1680 and early June of that year, the (circumpolar) Cas A supernova 
would have had its upper culmination in the daytime sky in the Northern hemisphere.
First observation of the supernova would almost certainly have been with naked eye, despite the
general use of the telescope\cite{footnote3} by late seventeenth-century astronomers.  
A Type Ic supernova at the distance of Cas A would have been unobservable by naked eye in 
daytime even at peak brightness.  The threshold 
visual magnitude for observation of a star by a human observer in the presence of 
Rayleigh-scattered sunlight may be estimated as described by Hughes\cite{Hughes1983}.  The minimum 
detectable brightness contrast in $lux$ detectable by human observers\cite{KnollEtAl1946,Hecht1947} is
converted to stellar magnitude with Russell's\cite{Russell1916,Russell1917} value for the stellar 
equivalent 1 $lux = -14.18$ visual magnitudes.  The atmospheric radiative transfer code 
MODTRAN4\cite{BerkEtAl1999,KniezysEtAl1996} was used to calculate the brightness of the sunlit sky 
at the February 14, 1680 position of the expansion center of Cas A as viewed from Greenwich 
for Julian dates
1680 February 2, February 12,  February 22/23, and March 3,
using Solar positions from
Gingerich and Welther$^{28}$.  Threshold $m_{V}$ magnitudes corresponding 
to a 
just-detectable star at these dates for Cas A altitudes near upper culmination appear in Table I.
The $m_{V}$ values in the table are probably in error by no more than $\pm 0.1$ magnitudes.  
Theshold $m_{V}$ values are higher for very low Solar altitudes, samples of which also appear in 
the table, but the plane-parallel MODTRAN model
is probably not to be relied upon for solar altitudes as low as $5^{\circ}$.

\begin{table}[ht]
Table 1.  Threshold V Magnitudes at location of Cas A 
{ \small \begin{tabbing}
date.....>>>.........\=.......Cas A altitude..\=......solar altitude..\=threshold $m_{V}$ \kill
date\>Cas A altitude\>solar altitude\>threshold $m_{V}$ \\
\rule{120mm}{0.1mm} \\
1680 Feb 2  \> $83.66^{\circ}\,$ \>  $22.62^{\circ}$ \> \,$-2.4$ \\
\;\;\;\;" \> $\,65.22^{\circ}\,$\> $\;5.75^{\circ}$\,\> $-1.0$ \\
1680  Feb 12 \> $83.40^{\circ}$ \> $28.64^{\circ}$ \> \,$-3.1$ \\
\;\;\;\;" \> $56.65^{\circ}$ \> $\;5.04^{\circ}$ \> \,$-0.8$ \\
1680 Feb 22 \> $51.78^{\circ}$ \> $\;8.07^{\circ}$ \> \,$-1.3$  \\
1680 Feb 23 \> $83.76^{\circ}$ \> $33.14^{\circ}$ \> \,$-2.8$ \\
1680  Mar 3  \>  $82.57^{\circ}$ \> $36.66^{\circ}$ \> \,$-3.0$ \\
\;\;\;\;" \> $42.71^{\circ}\,$ \> $\;5.51^{\circ}\,$ \> $-1.2$\,\\\end{tabbing}}
\end{table}

\begin{table}[ht]
Table 2.  Apparent V Magnitudes of SN1994I and SN1993J at distance of Cas A 
{ \small \begin{tabbing}
template...\=...$A_{v}$...\=peak/$m_{V}$..\=$m_{V}(170 d)$..\=$m_{V}(185 d)$..\=$m_{V}(185 d)$..
\=$m_{V}(185 d)$..\=$m_{V}(185 d)$..\=$m_{V}(200 d)$  \kill
template\>$A_{v}$\>peak $m_{V}$\>$m_{V}(170 d)$\>$m_{V}(185 d)$\>$m_{V}(200 d)$\>$m_{V}(220 d)$
\>$m_{V}(235 d)$\>$m_{V}(250 d)$ \\
\rule{160mm}{0.1mm} \\
SN1994I\>5.0 \> -0.43\> 5.0\> 5.3\> 6.0 \> -- \> -- \> -- \\
\;\;\;\;"\>5.3 \> -0.13\> 5.3\> 5.6\> 6.3 \> -- \> -- \> -- \\
\;\;\;\;"\>6.0 \>  \;0.57\> 5.5\> 5.8\> 6.5 \> -- \> -- \> -- \\
SN1993J\>5.0 \>  \;0.04\> -- \> --\> 4.6\> 5.0\> 5.2\>5.5\\
\;\;\;\;"\>5.3 \>  \;0.34\> -- \> -- \> 4.9\> 5.3\> 5.5\>5.8\\
\;\;\;\;"\>6.0 \>  \;1.04\> -- \> -- \> 5.5\> 6.0\> 6.2\>6.5\\
\end{tabbing}}
\end{table}

For purposes of discussion, consider SN1994I as a template for the Cas A 
outburst.  
Estimated apparent $V$ magnitudes for SN1994I 
at the distance of Cas A for maximum light 
and various times thereafter appear in Table II.~\cite{Richmond1996}  
If the values shown for $A_{v}$ are 
assumed uncertain by $\pm 0.45$ magnitudes, the apparent V magnitudes of Cas A in the table
should have a net uncertainty
$(-0.4,+0.5)$ from combined errors in extinction and distance modulus. The uncertainty arising 
solely from error in the distance  
modulus is $(-0.1,+0.2)$ magnitudes. Table II also contains apparent $V$ magnitudes for
SN1993J at the distance of Cas A.~\cite{Richmond1996b}
The SN1993J $V$ light curve is quite similar to SN1994I. 
While the peak 
$M_{V}$ of SN1993J is dimmer by about half a magnitude, its exponential tail is somewhat brighter
at the same time after maximum light,
and at the distance of Cas A, it attains the same visual magnitude as SN1994I as much as 60 days
later.  Without further adjustment of the overall brightness of the light curve, it is just possible
to fit it into the scenario for Cas A. 

Richmond et al. 
find a peak visual magnitude $M_{v}=-18.09^{+0.58}_{-0.58}$ for SN1994I.\cite{Richmond1996}
Taking $M_{v}=-18.09$
for the peak magnitude and $A_{v}=5$
the peak apparent visual magnitude of Cas A would have been $m_{v}(peak)=-0.43^{+0.2}_{-0.1}.$
In early February 1680,
the threshold visual magnitude which would have been observable by naked eye against the background
of Rayleigh-
scattered sunlight would have been brighter than $m_{v}(threshold)=-1$ for Solar altitudes above 
$5^{\circ}$.~\cite{footnote4}
If the Cas A explosion occurred between the new year and June, Cassiopeia would
have been been unobservable by naked eye anytime near upper culmination.       

The foregoing estimates show that, on this account, the position of the Cas A supernova in winter 
1680 would have been such that sunlight would have precluded naked-eye discovery anywhere near the 
meridian. This may go some ways toward accounting for the failure of contemporary astronomers to 
note any such phenomenon, but leaves unexplained why the circumpolar supernova was not observed 
at night.
Cas A would have been visible at peak brightness 
(whenever that occurred) during some portion of
nighttime at an appreciable altitude above the horizon.  On the second
of February, it should have exceeded the visible threshold shortly after sunset at an altitude 
of approximately $57^{\circ}$. At this altitude, atmospheric extinction would have been about 
0.1 V magnitudes, so that Cas A could have appeared as bright as 
$m_{V}=-0.3^{+0.2}_{-0.1}$ ($A_{v}=5$).

It is
necessary, then, to stipulate some reason for the absence of reported observations of the 
supernova during the winter
months apart from its unobservability during daylight.  Here the story, unavoidably, becomes 
speculative.
Apart from known factors such as the small number of astronomers active in the 
seventeenth century and the lack of any network 
for collection, dissemination, or archival of reports by non-astronomers, one might adduce the
discouraging effect on observation of winter conditions in the Northern hemisphere, including the 
possibility of extended periods of cloudy weather. 
The supernova could have been dimmer at its peak, or its extinction greater, and it could have 
decayed somewhat more slowly, than assumed to this point:  
On February second, an outburst with peak $M_{V} \ge -17$ or extinction $A_{v} \ge 7$ would 
have appeared 
to an observer 
with $m_{V}(peak) \ge 1.7^{+0.2}_{-0.1}$, in the range which van den Bergh
and Dodd suggest might have escaped notice by contemporary observers.~\cite{vdBDodd1970}

Return to the observation of 3 Cassiopeiae on the night of 1680 August 16$^{5}$.
Flamsteed reported stellar magnitudes as integer rank values, so that 
one may suppose that 3 Cassiopeiae could have been as bright as $m_{V}=5.5$.  The SN1994I light 
curve
could have faded to $m_{v}=5.5^{+0.2}_{-0.1}$ by day $170$ after maximum ($A_{v}=5$), while SN1993J 
could have done
so by day $200$ ($A_{v}=6$).  
Note that the rapid decay of SNeI is necessary in order to reach sixth magnitude within the requisite
period of 170-200 days after peak brightness; no normal SNII light curve is known to decay that 
rapidly.  

Although the particular scenario presented in this note requires the Cas A supernova
to have been observable at the time and location of the disputed report by Flamsteed, the explosion
could just as well have occurred mid to late winter of some other year in the late seventeenth 
century,  resulting an unprepossessing object of sixth magnitude at the upper culmination of 
Cassiopeia in August of that year.

\end{document}